# Understanding contact gating in Schottky barrier transistors from 2D channels

Abhijith Prakash[1, 2, §, *], Hesameddin Ilatikhameneh[1, 3, §], Peng Wu[1, 2] and Joerg Appenzeller[1, 2]

[1] School of Electrical and Computer Engineering, Purdue University, West Lafayette, Indiana 47907, USA.

[2] Birck Nanotechnology Center, Purdue University, West Lafayette, Indiana 47907, USA.

[3] Network for Computational Nanotechnology, 207 S. Martin Jischke Drive, West Lafayette, Indiana 47907, USA.

[§] These authors have contributed equally to this work.

[*] Address correspondence to prakash1@purdue.edu

**ABSTRACT**

In this article, a novel two-path model is proposed to quantitatively explain sub-threshold characteristics of back-gated Schottky barrier FETs (SB-FETs) from 2D channel materials. The model integrates the "conventional" model for SB-FETs with the phenomenon of contact gating – an effect that significantly affects the carrier injection from the source electrode in back-gated field effect transistors. The two-path model is validated by a careful comparison with experimental characteristics obtained from a large number of back-gated $WSe_2$ devices with various channel thicknesses. Our findings are believed to be of critical importance for the quantitative analysis of many three-terminal devices with ultrathin body channels.

Over the years, fabrication of back-gated (BG-) field-effect transistors (FETs) has become the most common way to build a three-terminal device on emerging materials to investigate their intrinsic properties and to understand the resulting carrier transport[1-17]. BG-FETs have been an attractive option particularly due to the ease of device fabrication and the resulting high yield. While often not employing a scaled dielectric, there have been numerous instances where a back-gating approach has been utilized for the initial demonstration of novel phenomena such as band-to-band tunneling, the impact of strain or observation of quantum oscillations in 2D systems, to just name a few[18-27]. What makes back-gated device structures special is that different from a conventional device layout, the entire channel segment underneath the source/drain contact region is under some influence of the gate. It is this particular behavior that needs to be understood in order for any quantitative device analysis to be relevant, which is the topic of this article.

Since chemical doping of low-dimensional materials is challenging and is still in its infancies, a transistor structure with highly doped source and drain regions connected to a gated channel, as employed for conventional metal-oxide semiconductor (MOS) FETs, is not common for exploratory devices. In fact, source and drain metal contacts are typically directly deposited onto the novel channel material, in this way only making direct contact to the very top. Such a structure when gated is commonly referred to as Schottky barrier (SB)-FET. Frequently, this top-contact design is combined with the use of a heavily doped substrate (*e.g.* silicon) isolated from the channel through a dielectric (*e.g.* silicon dioxide) as a large area gate of the device test structure, thus bringing the entire channel, including the source-to-channel and the drain-to-channel region under the gate control. Analyzing this type of structure has been the focus of many research articles and the description of SB-FETs in terms of a gated channel that is connected to a fixed barrier at the metal-to-channel interface (the Schottky barrier) has been successfully employed for a number of model systems including 1D channels like Si nanowires, carbon nanotubes and 2D channels like black phosphorus, $MoS_2$ and alike[28-32].

In this article, we will discuss in how far the "conventional" Schottky barrier model[31, 32] needs to be extended in general to include contact gating, an effect that had been discussed by us in 2009 in the context of graphene devices[33], to accurately describe the sub-threshold device characteristics from most two-dimensional (2D) materials. In particular, we propose here a general, physics-based parameter-free model to describe the electrical characteristics of back-gated SB-FETs with 2D channels, and demonstrate its validity by employing it to successfully explain the experimentally obtained characteristics of back-gated WSe$_2$ SB-FETs for various channel thicknesses.

**RESULTS AND DISCUSSION**

Any current I$_D$ in an SB-FET can be associated with either: (i) thermal current from purely thermionic carrier injection over the Schottky barrier or (ii) Schottky barrier current due to thermally assisted tunneling of charge carriers through the Schottky barrier. The conventional SB-FET model describes the sub-threshold region (OFF state) of the transfer characteristic (I$_D$-V$_{GS}$) with the help of a single equation, using Landauer formalism, assuming that the gate's control only extends over the channel (*i.e.* without including the segments underneath the source and drain contacts). As per the conventional SB-FET model[31, 32], the source-injected electron current per unit channel width is given by

$$I_D = \frac{2q}{h} \int_{E_C}^{\infty} M(E)\, T(E) f(E)\, dE \qquad (1)$$

where M(E) is the number of modes per unit width given by

$$M(E) = \frac{2}{h}\sqrt{2m_e(E - E_C)} \qquad (2)$$

For E < $\Phi_n$, T(E) is the probability of transmission through the Schottky contact as calculated by WKB approximation and is given by

$$T(E) = \exp\left(-\frac{8\pi}{3h}\sqrt{2m_e(\Phi_n - E)^3}\frac{\lambda}{(\Phi_n - E_C)}\right) \qquad (3)$$

For energies greater than the Schottky barrier height for electrons ($\Phi_n$), the probability of transmission is unity as this corresponds to pure thermal injection.

In the above equations, E is the electron energy with respect to the metal Fermi level at the source, f(E) is the Fermi function at the source given by $f(E) = [1 + \exp\left(\frac{E}{k_B T}\right)]^{-1}$, $m_e$ is the effective tunneling mass for electrons which is usually expressed as a multiple of the free-electron mass $m_0$, and $E_C$ is the gate-bias controlled conduction band minimum in the channel. The gate voltage at which $E_C = \Phi_n$ is known as the flat-band voltage ($V_{FB}$), which separates the thermal injection dominated gate voltage range from the Schottky barrier dominated one. In fact, $I_D$ can be divided into two components $I_{Ch-B}$ and $I_{SB-T}$ (*i.e.*, $I_D = I_{Ch-B} + I_{SB-T}$) where $I_{Ch-B}$ is due to thermal injection, limited by the channel potential below flat-band and by the Schottky barrier above flat-band. $I_{SB-T}$ is the additional current injected by tunneling through the Schottky barrier above flat-band (figure 1(a)). Since $I_{Ch-B}$ flows through the channel even if there is no tunneling through the Schottky barrier, it is regarded as the basic channel current.

λ is the characteristic length scale which defines the distance over which the potential changes from the metal-semiconductor interface to the channel. Several equations have been proposed in the literature for λ in an ultrathin-body channel[34-36], the two prominent ones being: (i) a square root scaling length given by $\lambda_S = \sqrt{\frac{\varepsilon_{body-x}}{\varepsilon_{ox}} t_{body} t_{ox}}$ and (ii) a generalized scaling length $\lambda_T$, the value of which is obtained by solving the equation $\frac{1}{\varepsilon_{ox}} \tan\left[\frac{2t_{ox}}{\lambda_T}\right] + \frac{1}{\varepsilon_{body-x}} \tan\left[\frac{2t_{body}}{\lambda_T}\right] = 0$. In the above expressions, $t_{ox}$ is the thickness of gate oxide, $\varepsilon_{ox}$ denotes the dielectric constant of the gate oxide, $\varepsilon_{body-x}$ refers to the in-plane dielectric constant of the channel material and $t_{body}$ is the body thickness of the ultrathin channel.

If the band movement in the channel is not controlled by the gate voltage ($V_{GS}$) in a one-to-one fashion, the entire $I_D$-$V_{GS}$ curve resulting from the conventional SB-FET model is "stretched" along the $V_{GS}$-axis by a factor γ (band movement factor) which is the ratio of the change in gate voltage to the change in actual channel potential, thereby deteriorating the inverse sub-threshold slope (SS = $d(V_{GS}) / d(\log(I_D))$) for both, the thermal and the SB dominated part of the characteristics. This implies that in the case of thermal injection dominated currents, SS would deviate from its ideal value of 60 mV/dec at room temperature, becoming 60γ mV/dec and in the case of Schottky barrier currents, SS, which is always larger than 60mV/dec[29, 30, 37-39], will further increase by the same factor γ.

**Necessity of a new model**

To test the validity of a model, benchmarking with experimental results is necessary. For such a comparison in the case of back-gated Schottky barrier transistors with 2D channels, a 2D material which exhibits a prominent Schottky barrier current branch as well as a thermal branch observable above the measurement noise floor, needs to be chosen. WSe$_2$, which is an important member of the family of two-dimensional transition metal dichalcogenides (TMDs)[18, 24, 40-45] is known to satisfy these requirements[18, 46].

In order to fabricate back-gated WSe$_2$ SB-FETs, flakes of WSe$_2$ were micro-mechanically exfoliated on top of substrates with 90nm SiO$_2$ thermally grown on highly doped silicon. Flakes of various thicknesses were identified by means of optical contrast after proper calibration and atomic force microscopy (AFM) in tapping mode. Electron beam lithography followed by electron beam evaporation was used to define source and drain contacts, each designed to have a contact length ($L_{contact}$) of 500nm. Ni was used as the contact metal. The channel lengths for all the devices were designed to be 1.5μm and the highly doped Si was used as the back-gate electrode. A schematic of the device structure is shown in figure 1(b). All electrical measurements were carried out at room-temperature at a vacuum of ~$10^{-6}$ Torr in a Lake Shore probe station using an Agilent semiconductor parameter analyzer.

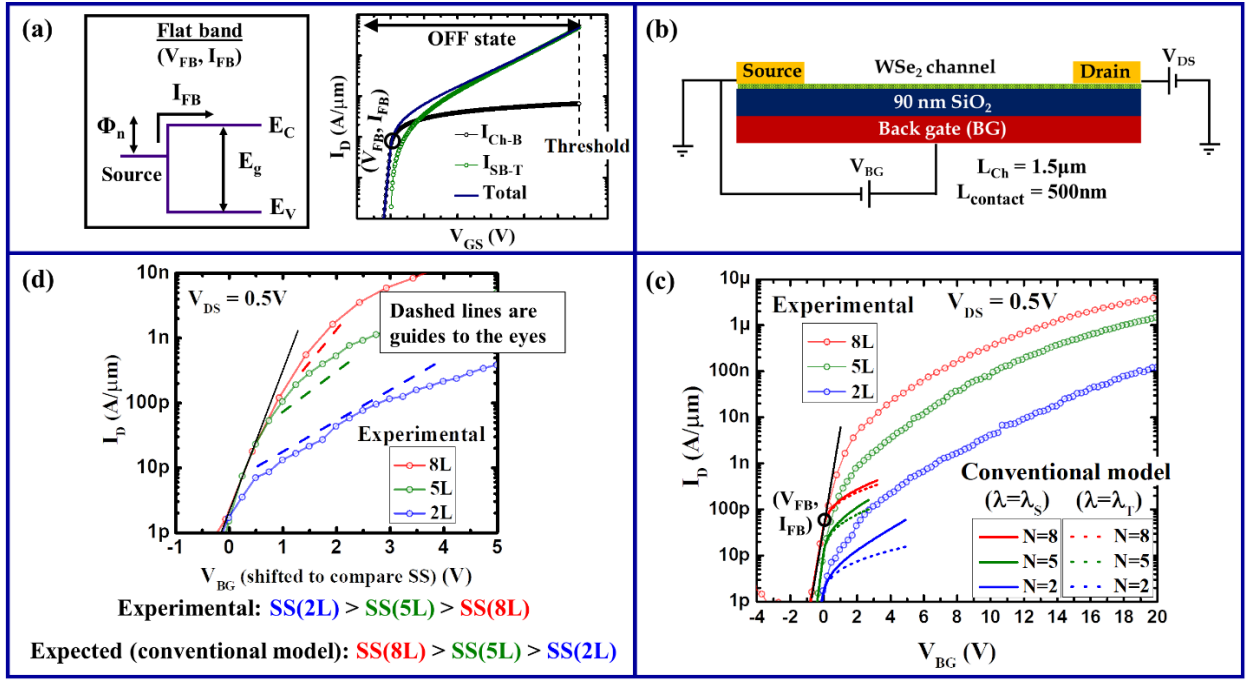

**Figure 1.** (a) Various components of current in the conventional SB-FET model. (b) Schematics of a back-gated WSe$_2$ Schottky barrier FET. (c) Comparison of the experimental device characteristics with simulations based on the conventional SB-FET model. (d) Comparison of SS for the same set of experimental transfer characteristics as in (c).

For each device (except the ones with single layer channels) the flat-band voltage ($V_{FB}$) was determined by carefully identifying the point of deviation from the thermal branch which is the point where $I_D$ deviates from its exponential dependence on $V_{BG}$ in the lowest current range (see figure 1(c)). From the corresponding current $I_{FB}$, the Schottky barrier height $\Phi_n$ was extracted using the equation $I_{FB} = \frac{2q}{h}\int_{\Phi_n}^{\infty} M(E)f(E)dE$, which is nothing but equation (1) at flat-band, by using an electron effective mass of $0.36m_0$, a value that is in accord with what has been reported in the literature[44, 47, 48]. All Schottky barrier heights extracted in this way ranged between 0.4eV to 0.5eV, depending on the body thickness as will be further discussed later. Since all measurements were performed at a drain bias of 0.5V which is greater than the Schottky barrier height, the drain side Schottky contact impact is eliminated[1, 32].

The value of γ for each device was determined experimentally by comparing the inverse sub-threshold slope (SS) of its thermal branch with 60γmV/dec. Channel thickness dependent values of the dielectric constant were obtained with the help of values reported in the literature[49] (see supplementary information I) by assuming $t_{body}$ to be 0.7nm times the number of WSe$_2$ layers (N) in the channel[42, 50].

Utilizing the extracted Schottky barrier heights from above, we employed the conventional SB-FET model, with both expressions - $\lambda_S$ and $\lambda_T$ - for $\lambda$, to explain our experimental results. Figure 1(c) illustrates the discrepancy between experimental data and the simulations. Not only does the thermal current transition at $V_{FB}$ into a Schottky barrier dominated current that is too low, but more importantly the gate voltage at which the conventional model predicts the device characteristics to transition into their ON-state (the $V_{BG}$-values at which the simulated curves end) is not even remotely close to where currents start to flatten out in the experimental curves which is for $V_{BG}$ ~ 15V to 30V. Attempts to artificially adjust parameters to achieve a better match between the conventionally modeled electrical response and the experimental data in terms of current levels requires much smaller Schottky barrier heights than those extracted from the flat band currents. However, these values are unrealistic considering the ambipolar nature of the experimental transfer characteristics (see supplementary information II) combined with the values of bandgaps previously extracted by us[37]. Moreover, even artificially correcting the current levels does still not yield an overall better fit (see supplementary information III in this context). Similarly, artificially varying $\varepsilon_{body-x}$ to its minimum possible value was also explored to achieve a fit with the conventional SB-FET model, but without any success. One of the most important discrepancies can be seen in figure 1(d), which shows that in addition to the other above arguments the experimental trend in SS with respect to body thickness and $\varepsilon_{body-x}$ is opposite to that predicted by the conventional model. A smaller body thickness should decrease the scaling length through both, a decrease in $t_{body}$ AND a decrease in $\varepsilon_{body-x}$ (see figure S1). All of the above implies that a major aspect in the description of the behavior of back-gated $WSe_2$ Schottky barrier FETs is missing in the conventional SB model.

**Importance of gate geometry**

The failure of the conventional SB-FET model in the domain of back gated 2D transistors, considering its success in modeling top gated transistors on 2D channels such as ultrathin body Si,[51] brings up the question: "Is there a fundamental difference between these two structures?" Since the conventional SB-FET model treats a top gate and a back gate identically, comparing top and bottom gated devices allows identifying their different impact on the channel. For that, we fabricated top gates on previously characterized back-gated devices covering the entire channel region in-between the source and drain contacts with 12nm thick $Al_2O_3$ using atomic layer deposition (ALD) and employing electron beam lithography plus electron beam evaporation to fabricate the top gates. Ni was used as the top gate metal. The resulting device structure, along with the corresponding SEM image, is shown in figure 2(a) and device characteristics for several $V_{BG}$ conditions while sweeping the top gate voltage $V_{TG}$ are displayed in figure 2(b).

If the two gates' impact on the channel is identical, changing the fixed voltage applied to one gate should result only in a threshold voltage shift in the transfer characteristics when the voltage applied to the other gate is swept. This is clearly not the case in figure 2(b) where the achievable ON-state current is a strong function of $V_{BG}$, which implies that carrier injection is ultimately limited by the back gate. This observation is in accord with the experimental results reported by H.C.P. Movva, et.al.[52], considering that the top gate and back gate are reversed in their device structure. As it is evident from figure 2(b), the top gate can only turn the device OFF, *i.e.*, it can only block the current. It can however not increase the current beyond a certain point by itself. This implies that the back gate can impact the channel region in portions not accessible to the top gate, which are the TMD segments right underneath the source and drain contacts. Since the back gate impact is substantial enough to modify the ON-state current levels by orders of magnitude, the conventional Schottky barrier model requires including these particular regions in the calculations of device characteristics explicitly which is the topic of the next section.

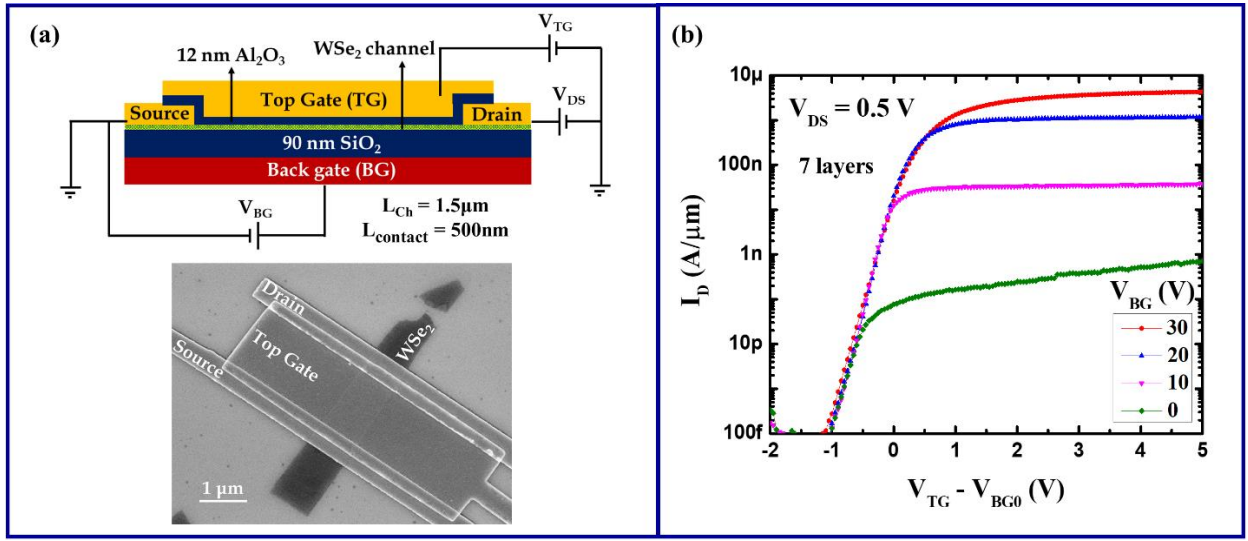

**Figure 2. (a) Modified device structure after the fabrication of a top gate along with the corresponding SEM image. (b) Top-gated transfer characteristics of a representative device for different values of $V_{BG}$ after compensating for the back gate induced threshold shifts $V_{BG0}$.**

**A new two-path model for back-gated Schottky barrier field-effect transistors**

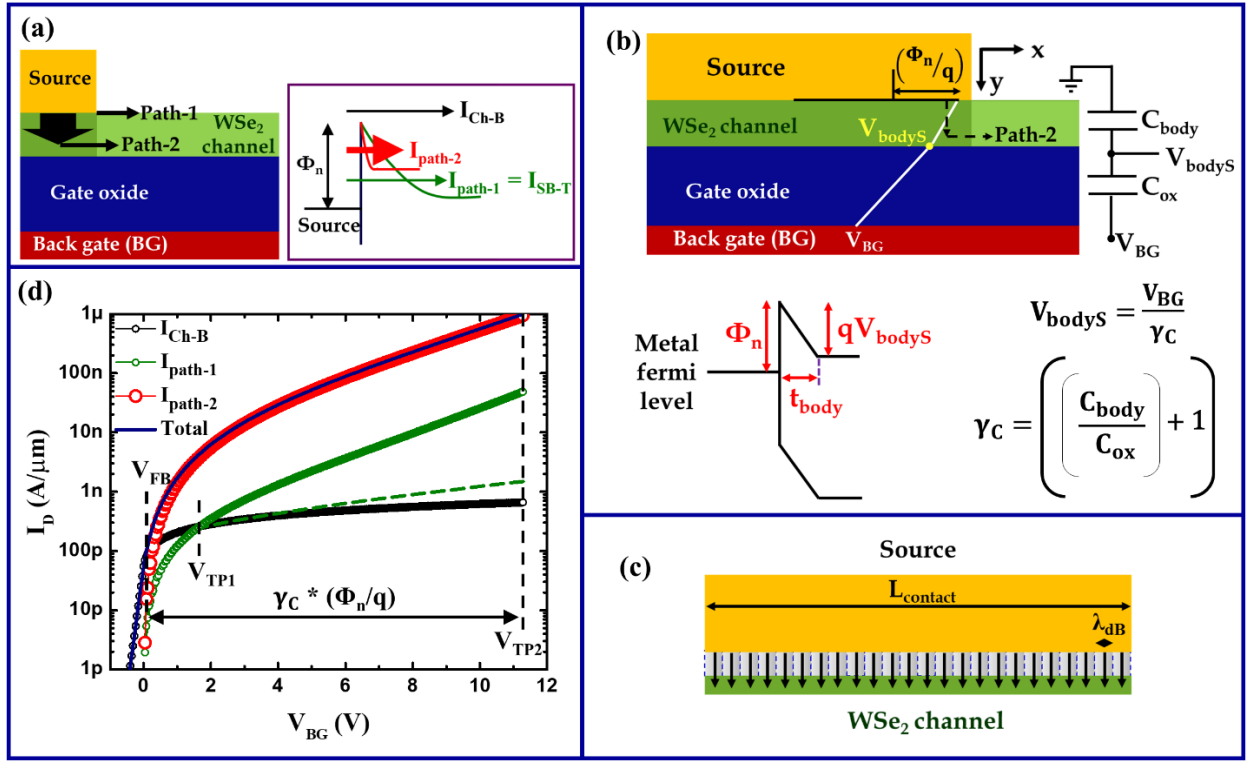

**Figure 3.** Illustration of the two-path model for back-gated SB-FETs where (a) shows the two injection paths, (b) explains diagrammatically the injection via path-2 and (c) presents a pictorial representation of the number of injecting states $N_i$ along the contact length. Shown in (d) is a typical transfer characteristic of a back-gated WSe$_2$ SB-FET, along with the individual contributions of each of the two paths, calculated as per the new model for a Schottky barrier height of 0.4eV and a body thickness of 7nm by assuming a square root scaling length $\lambda_S$ for path-1. Green circles assume continuous band movement for path-1 even above its threshold ($V_{TP1}$) whereas green dashed lines assume slowed down band movement for path-1 above threshold as described in the text.

In order to account for the aforementioned "additional" effect of a back gate in the contact region, we are proposing here a so called "two-path" model (see figure 3(a)). In this model, similar to the conventional model, the total current below flat-band is limited to the basic channel current $I_{Ch-B}$ since the channel resistance, by virtue of its barrier height, dominates the total resistance in this regime. Beyond flat-band, apart from allowing $I_{Ch-B}$, the back gate has two separate functions: (i) The back gate modulates the carrier injection via Schottky barrier tunneling right at the edge of the source-to-channel region as in the conventional SB-model (path-1) and (ii) allows simultaneously for injection into deep-lying layers of the TMD flake due to the electric field that is built up by $V_{BG}$ underneath the source contact (path-2). The sum of all these currents is the $V_{BG}$-dependent total current through the entire device characteristics.

In order to model path-2 for carrier injection, it is important to examine the potential profile in the channel region underneath the contacts in a back-gated device. Since the gate voltage drops across two dielectrics, the semiconducting channel material and the back oxide, a simple capacitance divider, as shown in figure 3(b), that treats the portion of the device underneath the source as a series arrangement of two parallel plate capacitors $C_{body}$ and $C_{ox}$ can be employed. Here $C_{body} = \varepsilon_{body-y}/t_{body}$ and $C_{ox} = \varepsilon_{ox}/t_{ox}$, where $\varepsilon_{body-y}$ and $\varepsilon_{ox}$ are the respective permittivities of the channel material and the oxide in y (out-of-plane) direction, $t_{body}$ and $t_{ox}$ are the respective thicknesses of the channel body and the gate oxide. Since we are dealing with the device's OFF-state, the carrier density in the channel underneath the source contact is small and hence the potential profile along the thickness of the channel (y-direction) is almost linear. The total potential drop $V_{bodyS}$ across the channel body under the source, in the y-direction, can be obtained by solving the above-described capacitance network to be $V_{BG}/\gamma_C$ where $\gamma_C$ is the band movement factor underneath the contact given by:

$$\gamma_C = \left( \left( \frac{t_{ox} * \varepsilon_{body-y}}{t_{body} * \varepsilon_{ox}} \right) + 1 \right) \qquad (4)$$

Figure 3(b) shows the potential profile along the semiconducting channel underneath the source contact. Carrier injection along path-2 depends on the vertical electric field $V_{bodyS}/t_{body}$. We model this current as a tunneling current through a triangular barrier with the barrier height being equal to the Schottky barrier height $\Phi_n$ and the tunneling distance given by the channel thickness $t_{body}$ as shown in figure 3(b). Accordingly, the current per unit channel width for path-2 can be written as

$$I_{Path2} = \frac{2q}{h} \int_{E_{CS}}^{\Phi_n} N_i(E)\, M_S(E)\, T_{WKB-S}(E) f(E) dE \qquad (5)$$

where $E_{CS} = \Phi_n - qV_{bodyS}$, is the conduction band minimum at the bottom of the channel body under the source contact, $M_S(E)$ captures the number of 2D modes per unit width and $T_{WKB-S}(E)$ is the probability of transmission through the triangular barrier along path-2 in WKB approximation. $M_S(E)$ and $T_{WKB-S}(E)$ are given by equations (2) and (3) respectively when $E_C$ is replaced by $E_{CS}$, and $\lambda$ is replaced by $t_{body}$. $f(E)$ is the Fermi function at the source and $N_i(E)$ is the number of injecting states along the contact length $L_{contact}$, which is given by

$$N_i(E) = \frac{L_{contact}}{h} \sqrt{2m_e(E - E_{CS})} \qquad (6)$$

To obtain the above expression for $N_i(E)$, we have assumed that the potential drop across the channel body, which is responsible for the carrier injection, is identical over the entire contact area $A_C$ ($A_C$ = device width*$L_{contact}$). To calculate the current per unit width at any energy, the number of 2D modes per unit width $M_S(E)$ has to be multiplied by the number of injecting states $N_i(E)$ along the contact length $L_{contact}$ (see supplementary note in this context). Since each injecting state "occupies" a length segment equal to the de-Broglie wavelength of an electron in the semiconductor (figure 3(c)) *i.e.*, $2\pi/k$ where k is the magnitude of the wave vector, the total number of injecting states $N_i$ along $L_{contact}$ is equal to $\frac{L_{contact}}{(2\pi/k)}$, which results in the expression presented in equation (6) when a parabolic energy dispersion in the semiconductor is assumed. When the back-gate voltage $V_{BG}$ is varied, $V_{bodyS}$ changes as $V_{BG}/\gamma_C$, $E_{CS}$ changes as $\Phi_n - qV_{bodyS}$. Then $N_i$ is calculated for every $E-E_{CS}$ as per equation (6), and used in equation (5) to obtain $I_{path2}$.

Figure 3(d) shows simulated transfer characteristic of a back-gated WSe$_2$ SB-FET, along with the individual contributions of both the injection paths, calculated for a Schottky barrier height of 0.4eV and a body thickness of 7nm by assuming a square root scaling length $\lambda_S$ for path-1. $V_{TP1}$ and $V_{TP2}$ in the figure refer to the threshold voltages of path-1 and path-2 respectively, where "threshold voltage" refers to the voltage at which the conduction band edge in the corresponding path gets aligned with the source Fermi level. $I_{path-1}$ shown in the figure was calculated by assuming that the band movement for path-1 continues one-to-one with $V_{BG}/\gamma$ even above its threshold $V_{TP1}$. The consequence of this assumption is that when $V_{BG} = V_{TP2}$, the conduction band in the conventional channel would be ~1.8eV below the valance band edge at the source metal-to-semiconductor contact interface. Since this is a highly unrealistic situation, we have shown by the dashed green line, the case where the band movement slows down after $V_{TP1}$ is reached and moves such that the conduction band in the channel reaches the valance band edge at the source metal-to-semiconductor contact interface when $V_{BG} = V_{TP2}$. Since in both the cases, the contribution of $I_{path-1}$ to the total current is negligible, assumptions regarding the band movement for path-1 above $V_{TP1}$ do not make a considerable difference under the circumstances considered here. While calculating $I_{path-1}$ for channel potentials exceeding 0.5V above $V_{FB}$, though the impact of the drain side Schottky barrier has been considered, it was found to have a negligible impact for the large $V_{DS}$-value of 0.5V considered here. It is important to realize that there is a significant difference in the electrostatic gate control of the potentials underneath the contact and in the conventional channel. Under the contact, the ratio of $C_{body}$ and $C_{ox}$ determines the band movement factor $\gamma_C$[53], resulting in a body-thickness and material dependent gate control whereas in the conventional channel, γ and hence the gate control is body-thickness independent and much stronger. As a result, path-1 reaches its threshold voltage $V_{TP1}$ at a much smaller gate bias compared to path-2 $V_{TP2}$ as shown in the figure. Since currents above flat band due to path-2 are much larger than those due to path-1 in the present case (see supplementary section IV for a counter example), the threshold voltage visible in the full device characteristic is that of path-2 and the resulting stretch of the transfer characteristics is much larger compared to that due to path-1 (figure 3(d)). As the strengths of the back gate control (*i.e.*, ratios of change in channel potential to change in gate voltage) are different for the two paths, we have considered here an undoped channel that ensures that the band bending situations for path-1 and path-2 coincide at flat-band. Different band offsets might result from doping - intentional or unintentional - or from the work function difference between the top and bottom gates in case of double gated structures.

Simulations based on this two-path model match well with the transfer characteristics of all devices for various body thicknesses as shown in figure 4. In total more than 28 devices have been fabricated and the characteristics in figure 4 are good representations of all devices included in this study. It is important to note that apart from the Schottky barrier heights $\Phi_n$, only two parameters – the electron effective mass of $0.36m_0$ and the channel thickness dependent dielectric constant - were used as input parameters for the new model and both of those were taken from the literature [references [44,49] and supplementary information I]. Moreover, the Schottky barrier heights for electrons $\Phi_n$ obtained using the two-path model (figure 5) are in good agreement with previously reported values[32,46] considering that in these articles the bandgap was assumed to have a certain value. Though the simulated curves shown in figure 4 assume a square root scaling length $\lambda_S$ for path-1, employing the generalized scale length $\lambda_T$ does not make a considerable difference, since the contributions of path-1 to the current are negligible in the WSe$_2$ FETs as illustrated in figure 3(d). Also, for simulations, band movement for path-1 is assumed to slow down beyond its threshold $V_{TP1}$, though the impact of this assumption on the final curve is negligible as mentioned in the previous paragraph.

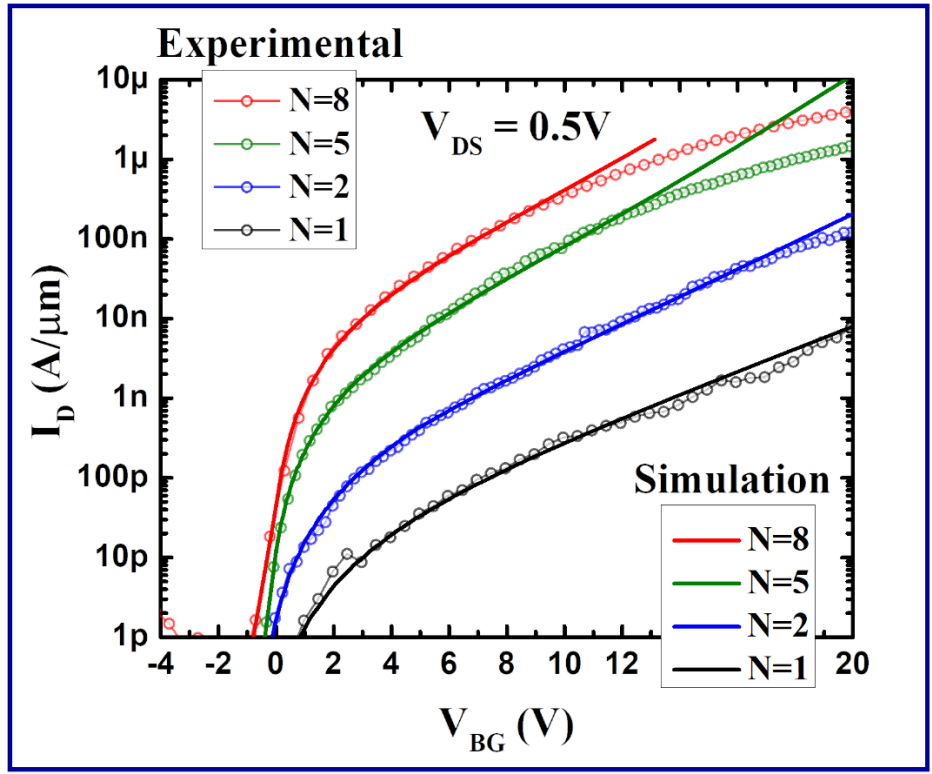

**Figure 4. Comparison between the experimental device characteristics obtained from various back-gated WSe$_2$ SB-FETs and simulations performed based on the new model.**

Deviation of the experimental curves from the simulated ones at high current levels are expected as the transport at such high currents involves a substantial number of injected charge carriers causing scattering in the channel - both underneath the contacts and in the conventional channel. The accumulation of carriers in or close to the device ON-state also implies a reduced gate response that is not captured by our model, which is valid only below threshold. In fact, to describe the ON-state performance of TMD devices a complicated interplay between mobility, carrier density, density of states in the channel, intra and inter-layer resistances and the gate controlled Schottky barriers need to be simultaneously taken into account[54-59], which is not the topic of this study.

Since the current contribution due to path-2 is proportional to $L_{contact}$ because of operation in the device OFF-state, it can be reduced by decreasing $L_{contact}$. Also, as mentioned before, the band movement for path-2 is much slower than that for path-1 and the relative strength of the gate control depends on the details of the material system and in particular the dielectric constants. Thus, for certain material systems and/or contact lengths the current injection via path-1 can turn out to be considerably higher than that via path-2 and the conventional Schottky barrier model is applicable. An example of this case that is closely related to our previously reported analysis of black phosphorus devices[31] is discussed in the supplementary information IV. Also, in 1D channels like nanotubes and nanowires one frequently finds device layouts where contacts encase the channel to a large extent and screening prevents the applicability of our model.

Last, we used the above insights into the electron Schottky barrier height $\Phi_n$ as a function of layer number in combination with our previous findings on the change of transport bandgap $E_g$ with body thickness for WSe$_2$[37] to determine the Schottky barrier height for hole injection $\Phi_p$ using the equation $E_g = \Phi_n + \Phi_p$. The Schottky barrier heights thus extracted are plotted in figure 5 as a function of flake thickness. It is apparent from figure 5 that while $\Phi_n$ changes by only ~100meV over the thickness range presented, most of the bandgap change occurs in accord with a change of $\Phi_p$.

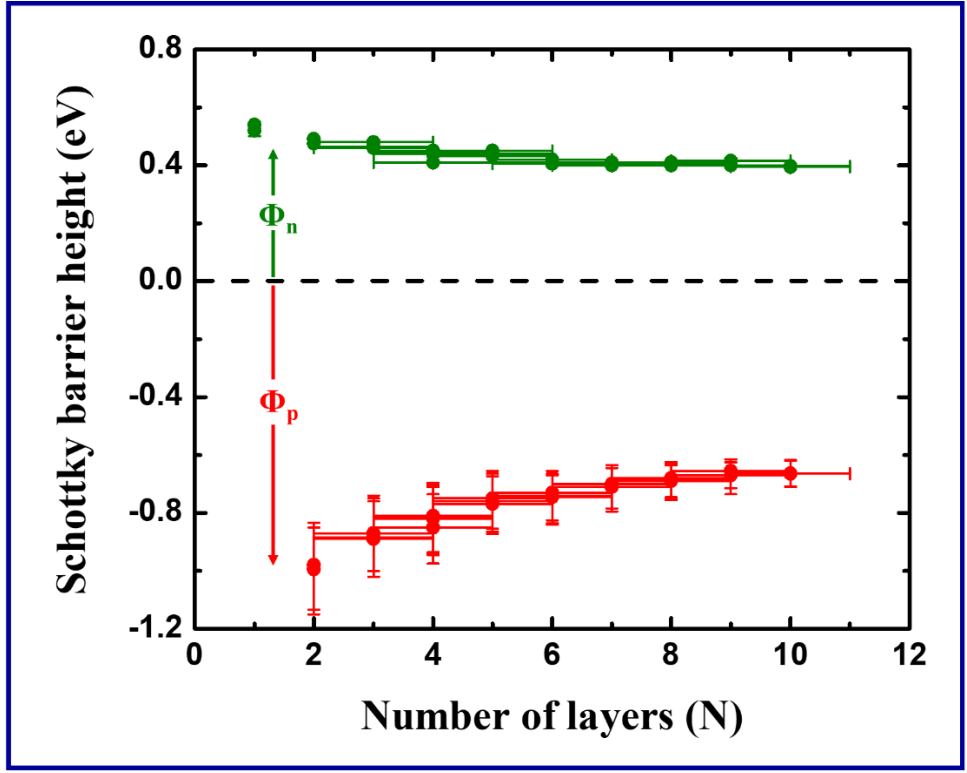

**Figure 5. Extracted Schottky barrier heights as a function of flake thickness for WSe$_2$ with Ni as the contact metal.**

**CONCLUSION**

In conclusion, we have proposed a comprehensive, physics-based model to describe the electrical response of back-gated Schottky barrier FETs with an ultrathin body channel by considering an additional current path for the first time. The new model was validated by means of comparison with a sizable amount of electrical characteristics from devices encompassing a wide range of channel thicknesses. Most importantly, in this study we have unveiled the significant role of the channel portion underneath the contacts in describing the carrier transport in transistors with 2D materials employed as channel materials.

**ACKNOWLEDGMENT**

This work was supported in part by the Center for Low Energy Systems Technology (LEAST), one of six centers supported by the STARnet phase of the Focus Center Research Program (FCRP), a Semiconductor Research Corporation program sponsored by MARCO and DARPA.

**AUTHOR CONTRIBUTIONS**

A.P. carried out the experimental work, identified the contact gating and vertical carrier injection based on the experimental results. H.I. proposed the preliminary model. A.P. improved the model with the help of J.A. and performed simulations. P.W. provided inputs to the simulation effort. J.A. directed the project. A.P., H.I. and J.A. co-wrote the manuscript. All the authors have read and approved the final manuscript.

## SUPPLEMENTARY INFORMATION

### I. Body thickness dependent dielectric constants in WSe$_2$

Similar to other 2D crystals, the dielectric constant, both in-plane (x-direction) and out-of-plane (y-direction), of WSe$_2$ depends on its thickness. Since the values of dielectric constants for WSe$_2$ have been reported in [1] only for a few body thicknesses, we have employed a spline interpolation approach to determine the dielectric constants in WSe$_2$ as a function of body thickness, with the dielectric constant values reported in [1] acting as fixed points. Figure S1(a) and figure S1(b) show the resulting in-plane dielectric constant values ($\varepsilon_{body-x}$) and out-of-plane dielectric constant values ($\varepsilon_{body-y}$) employed for the calculations in the main manuscript, respectively.

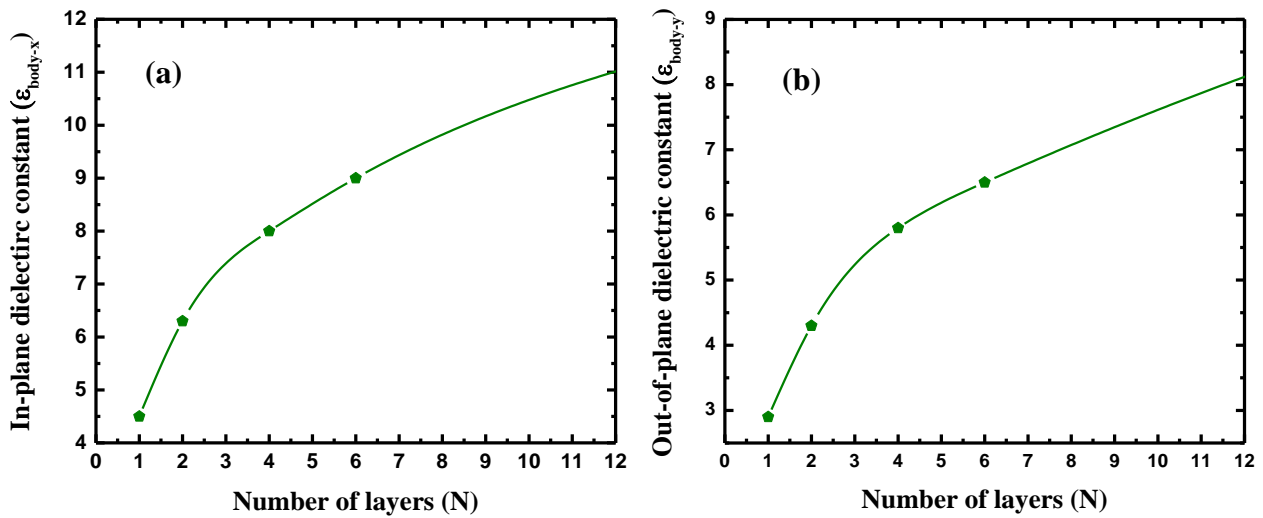

Figure S1: Spline-connected plots of (a) in-plane dielectric constant and (b) out-of-plane dielectric constant in WSe$_2$ as a function of its body thickness.

## II. Ambipolar device characteristics

Figure S2 presents the same set of experimental transfer characteristics as in figure 1(b) in the main text, with both the electron and hole current branches shown.

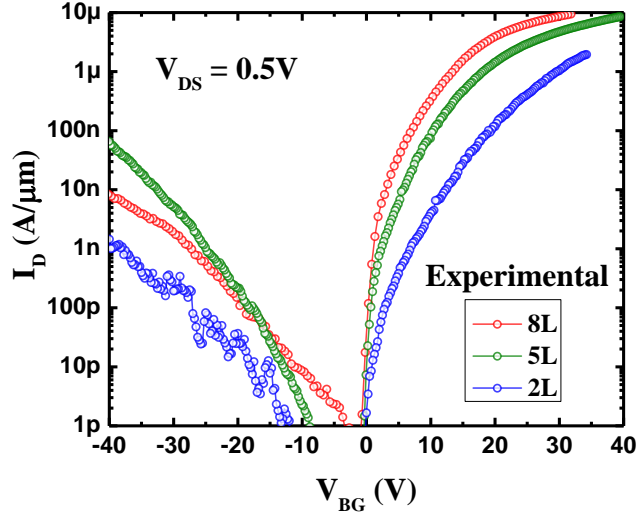

Figure S2: Representative set of ambipolar experimental device characteristics.

## III. Current level adjustment using the conventional SB-FET model

An attempt is made in figure S3 to check if artificially selecting a smaller Schottky barrier height, even if it is unrealistic, could help to describe the experimental device characteristics within the conventional SB-FET model. Figure S2 clearly shows that even under these assumptions the full experimental device characteristics are not captured by the model.

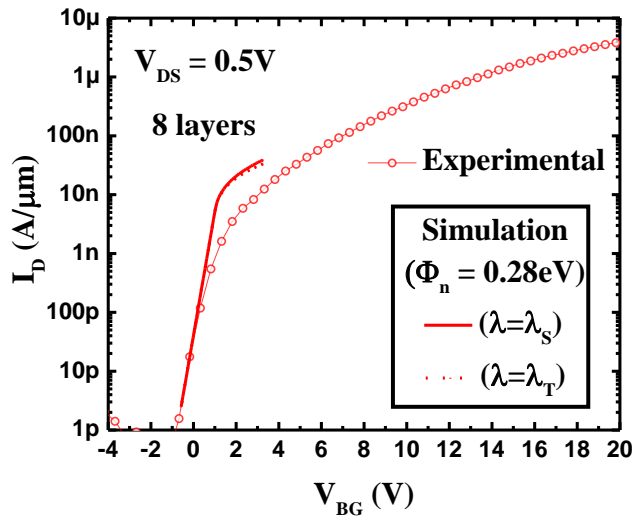

Figure S3: Comparison of the experimental transfer characteristic of an 8 layer WSe$_2$ device with simulations based on the conventional SB-FET model using an artificially selected Schottky barrier height of 0.28eV.

## IV. Illustration of a special case

Presented here is an example of a back-gated SB-FET where currents due to path-2 are less than those due to path-1 for a range of back-gate voltages ($V_{BG}$). For this example, the Schottky barrier height $\Phi_n$ has been chosen to be 0.2eV for a 5nm thick channel. An isotropic dielectric constant of 10 and an effective mass of $0.14m_0$ have been assumed for the channel material for a contact length $L_{contact}$ of 250nm. The resulting simulated electron currents for both paths discussed in the main text are shown in figure S4(a). Since currents predicted by the conventional SB-FET model are much larger than those due to path-2, the conventional SB-FET model can be used to explain the device characteristics fairly well (figure S4(b)). It should be noticed that the material parameters assumed in this example for electron transport match the corresponding parameters for hole transport in Black Phosphorus (BP) if anisotropic transport contributions in BP are ignored [2-5]. Thus, hole transport in BP, as described by Penumatcha, A. V. et. al. [3] is a good example of a 2D channel where the discussed contact gating effects are less pronounced in the resulting device characteristics. (The contact length assumed here is representative of the contact lengths designed in [3]).

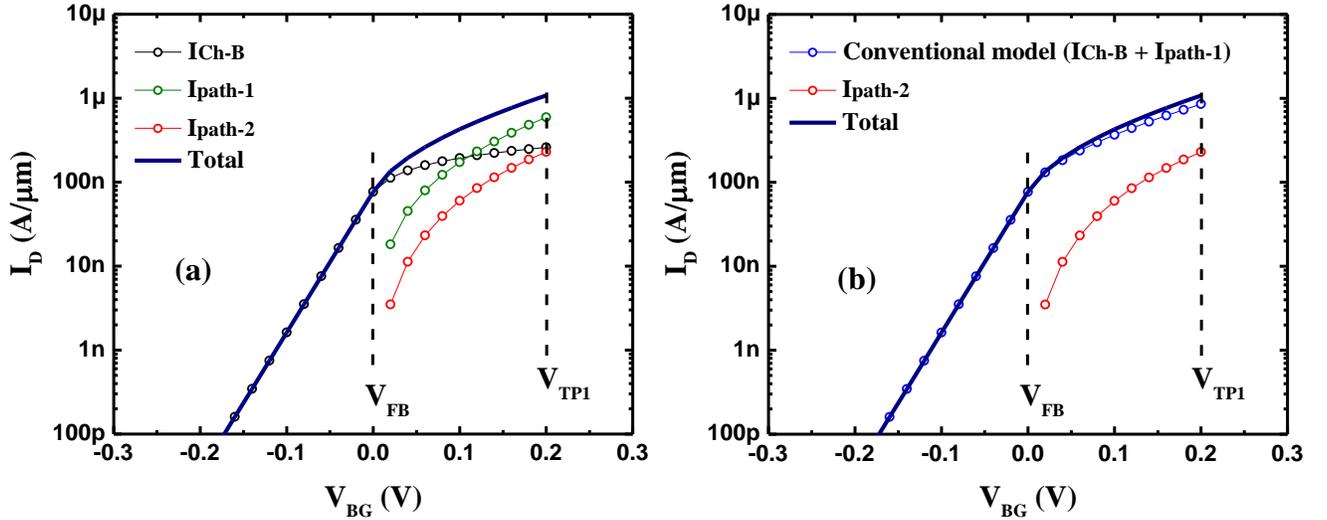

Figure S4: Illustration of a case where currents due to path-2 are significantly less than those predicted by the conventional SB-FET model, such that the conventional SB-FET model can be used to describe the device characteristics.

## V. AFM images

Shown below are the AFM images of a few representative WSe$_2$ flakes used in our study.

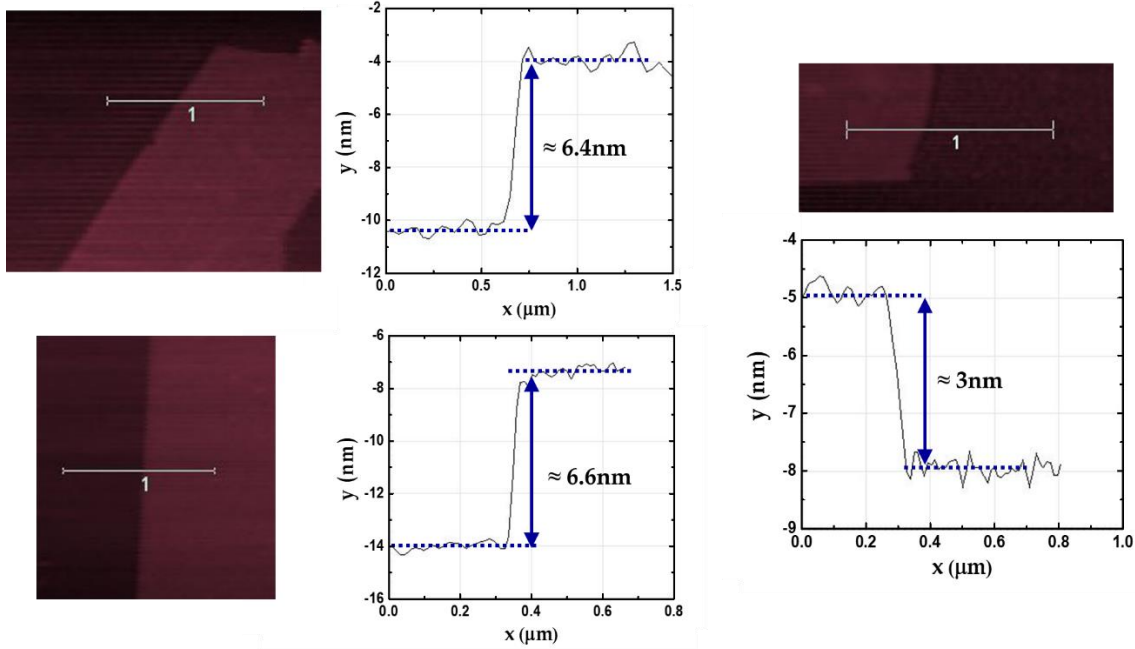

**Supplementary note:** The validity of I$_{path-2}$ being proportional to L$_{contact}$ through N$_i$(E) is limited to the device OFF state, where a finite carrier density underneath the contacts and in the channel can be ignored. In this gate voltage range, the scattering limited current in the device ON state never becomes the dominant resistance of the entire system, in which case the current scaling with the contact length would obviously not be applicable anymore. Moreover, for very large contact lengths, we expect that the linear dependence of N$_i$(E) on L$_{contact}$ will no longer hold true even in the OFF state, since the access resistance to the channel itself would become dominant, rather than the injection into the TMD.

**Supplementary References**

1. Kumar, A. & Ahluwalia, P. K. Tunable dielectric response of transition metals dichalcogenides MX$_2$ (M=Mo, W; X=S, Se, Te): effect of quantum confinement. *Phys. B* **407,** 4627-4634 (2012).

2. Morita, A. Semiconducting black phosphorus. *Appl. Phys. A* **39,** 227– 242 (1986).